\documentclass[11pt]{cernrep}
\usepackage{epsfig}
\usepackage{here}

\def\reffi#1{\mbox{Fig.~\ref{#1}}}

\def\refta#1{\mbox{Table~\ref{#1}}}
\def\reftas#1{\mbox{Tables~\ref{#1}}}
\def\refse#1{\mbox{Sect.~\ref{#1}}}
\def\refses#1{\mbox{Sects.~\ref{#1}}}

\def\citere#1{\mbox{Ref.~\cite{#1}}}

\newcommand{\GF}{\mathswitch {G_\mu}}
\newcommand{\sw}{\mathswitch {s_{\scriptstyle  W}}}

\def\mathswitch#1{\relax\ifmmode#1\else$#1$\fi}
\def\mathswitchr#1{\relax\ifmmode{\mathrm{#1}}\else$\mathrm{#1}$\fi}

\begin{document}

\title{\boldmath{PRECISION CALCULATIONS FOR ASSOCIATED 
$WH$ AND $ZH$ PRODUCTION AT HADRON COLLIDERS}}

\author{O.~Brein$^1$, M.~Ciccolini$^2$, S.~Dittmaier$^3$, A.~Djouadi$^4$, 
R.~Harlander$^5$ and M.~Kr\"amer$^2$}

\institute{%
${}^1$ Institut f\"ur Theoretische Physik, RWTH Aachen, Germany \\
${}^2$ School of Physics, The University~of~Edinburgh, Scotland \\
${}^3$ Max-Planck-Institut f\"ur Physik (Werner-Heisenberg-Institut),
M\"unchen, Germany \\
${}^4$ Laboratoire de Physique Math\'ematique et Th\'eorique, 
Universit\'e de Montpellier II, France \\
${}^5$ Institut f\"ur Theoretische Teilchenphysik, Universit\"at Karlsruhe, 
Germany}

\maketitle

\begin{abstract}
Recently the next-to-next-to-leading order QCD corrections and the  
electro\-weak ${\cal O}(\alpha)$ corrections to the Higgs-strahlung
processes $p\bar p/pp\to WH/ZH+X$ have been calculated. 
Both types of corrections are of the order of 5--10\%.
In this article the various corrections are briefly discussed and
combined into state-of-the-art predictions for the cross sections. 
The theoretical uncertainties from renormalization/factorization scales
and from the parton distribution functions are discussed.
\end{abstract}

\section{INTRODUCTION}

At the Tevatron, 
Higgs-boson production in association with $W$ or
$Z$~bosons, $p\bar p \to WH/ZH+X$,
is the most promising discovery channel for a SM Higgs
particle with a mass below about 135 GeV, where decays into
$b\bar{b}$ final states are dominant~\cite{Carena:2000yx,Stange:ya}.
At the $pp$ collider LHC other Higgs-production
mechanisms play the leading role \cite{atlas_cms_tdrs}, but nevertheless these
Higgs-strahlung processes should be observable.

At leading order (LO), the production of a Higgs boson in association
with a vector boson, $p\bar p \to VH+X, (V=W,Z)$ proceeds through
$q\bar{q}$ annihilation~\cite{Glashow:ab}, $q\bar{q}' \to V^* \to VH$.
The next-to-leading order (NLO) QCD corrections coincide with those to
the Drell-Yan process and increase the cross section by about
30\%~\cite{Han:1991ia}.  Beyond NLO, the QCD corrections to $VH$
production differ from those to the Drell-Yan process by contributions
where the Higgs boson couples to a heavy fermion loop.  The impact of
these additional terms is, however, expected to be small in
general~\cite{Dicus:1985wx}.  Moreover, for $ZH$ production the
one-loop-induced process $gg\to ZH$ contributes at
next-to-next-to-leading order (NNLO).  The NNLO corrections
corresponding to the Drell-Yan mechanism as well as the $gg\to ZH$
contribution have been calculated in \citere{Brein:2003wg}.  These
NNLO corrections further increase the cross section by the order of
5--10\%.  Most important, a successive reduction of the
renormalization and factorization scale dependence is observed when
going from LO to NLO to NNLO. The respective scale uncertainties are
about 20\% (10\%), 7\% (5\%), and 3\% (2\%) at the Tevatron (LHC).  At
this level of accuracy, electroweak corrections become significant and
need to be included to further improve the theoretical prediction.  In
\citere{Ciccolini:2003jy} the electroweak ${\cal O}(\alpha)$
corrections have been calculated; they turn out to be negative and
about --5\% or --10\% depending on whether the weak couplings are
derived from $G_\mu$ or $\alpha(M_Z^2)$, respectively.  In this paper
we summarize and combine the results of the NNLO corrections of
\citere{Brein:2003wg} and of the electroweak ${\cal O}(\alpha)$
corrections of \citere{Ciccolini:2003jy}.

The article is organized as follows.
In \refses{se:QCD} and \ref{se:EW} we describe the salient features
of the QCD and electroweak corrections, respectively.
Section~\ref{se:numres} contains explicit numerical results on the
corrected $WH$ and $ZH$ production cross sections, including a
brief discussion of the theoretical uncertainties originating from the
parton distribution functions (PDFs).
Our conclusions are given in \refse{se:concl}

\section{QCD CORRECTIONS}
\label{se:QCD}

The NNLO corrections, i.e.\ the contributions at ${\cal O}(\alpha_{\rm
  s}^2)$, to the Drell-Yan process $p\bar p/pp \to V^*+X$ consist of
the following set of radiative corrections:
\begin{itemize}
\item 
two-loop corrections to $q\bar{q} \to V^*$, which have to be
multiplied by the Born term,
\item 
one-loop corrections to the processes $qg \to qV^*$ and $q\bar{q}
\to gV^*$, which have to be multiplied by the tree-level $g q$ and $q\bar{q}$
terms,
\item 
tree-level contributions from $q\bar{q}, qq,qg, gg \to V^*+$ 2
partons in all possible ways; the sums of these diagrams for a given initial
and final  state have to be squared and added.
\end{itemize}
These corrections have been calculated a decade ago in 
\citere{Hamberg:1990np} and have recently been updated \cite{Harlander:2002wh}.
They represent a basic building block in the NNLO corrections to
$VH$ production. There are, however, two other sources of 
${\cal O}(\alpha_s^2)$ corrections:
\begin{itemize}
\item 
irreducible two-loop boxes for $q\bar{q}'\to VH$
where the Higgs boson couples via heavy-quark loops to two gluons that
are attached to the $q$ line,
\item 
the gluon--gluon-initiated mechanism $gg \to ZH$ \cite{Barger:1986jt}
at one loop;
it is mediated by closed quark loops which induce $ggZ$ and $ggZH$
couplings and contributes only to $ZH$ but not to $WH$ production.
\end{itemize}
In \citere{Brein:2003wg} the NNLO corrections to $VH$ production
have been calculated from the results \cite{Harlander:2002wh} on
Drell-Yan production and completed by the (recalculated) contribution
of $gg \to ZH$. The two-loop contributions with quark-loop-induced $ggZ$ 
or $ggH$ couplings are expected to be very small and have been neglected.

The impact of higher-order (HO) QCD corrections is usually quantified by
calculating the $K$-factor, which is defined as the ratio between the cross sections
for the process at HO (NLO or NNLO), with the value of $\alpha_{\rm s}$ and the PDFs
evaluated also at HO, and the cross section at LO,  with $\alpha_{\rm s}$
and the PDFs consistently  evaluated also at LO:
$K_{\rm HO}=\sigma_{\rm HO}(p\bar p/pp \to VH+X) / \sigma_{\rm LO}(p\bar p/pp\to VH+X)$.
A $K$-factor for the LO cross section, $K_{\rm LO}$, may also be
defined by  evaluating the latter at given factorization and renormalization
scales and normalizing to the LO cross sections evaluated at the central scale,
which, in our case, is given by $\mu_F=\mu_R=M_{VH}$, where
$M_{VH}$ is the invariant mass of the $VH$ system.

\begin{figure}
\begin{center}
{ \unitlength 1cm
\begin{picture}(15.5,6.0)
\put(-2.2,-5.7){\includegraphics{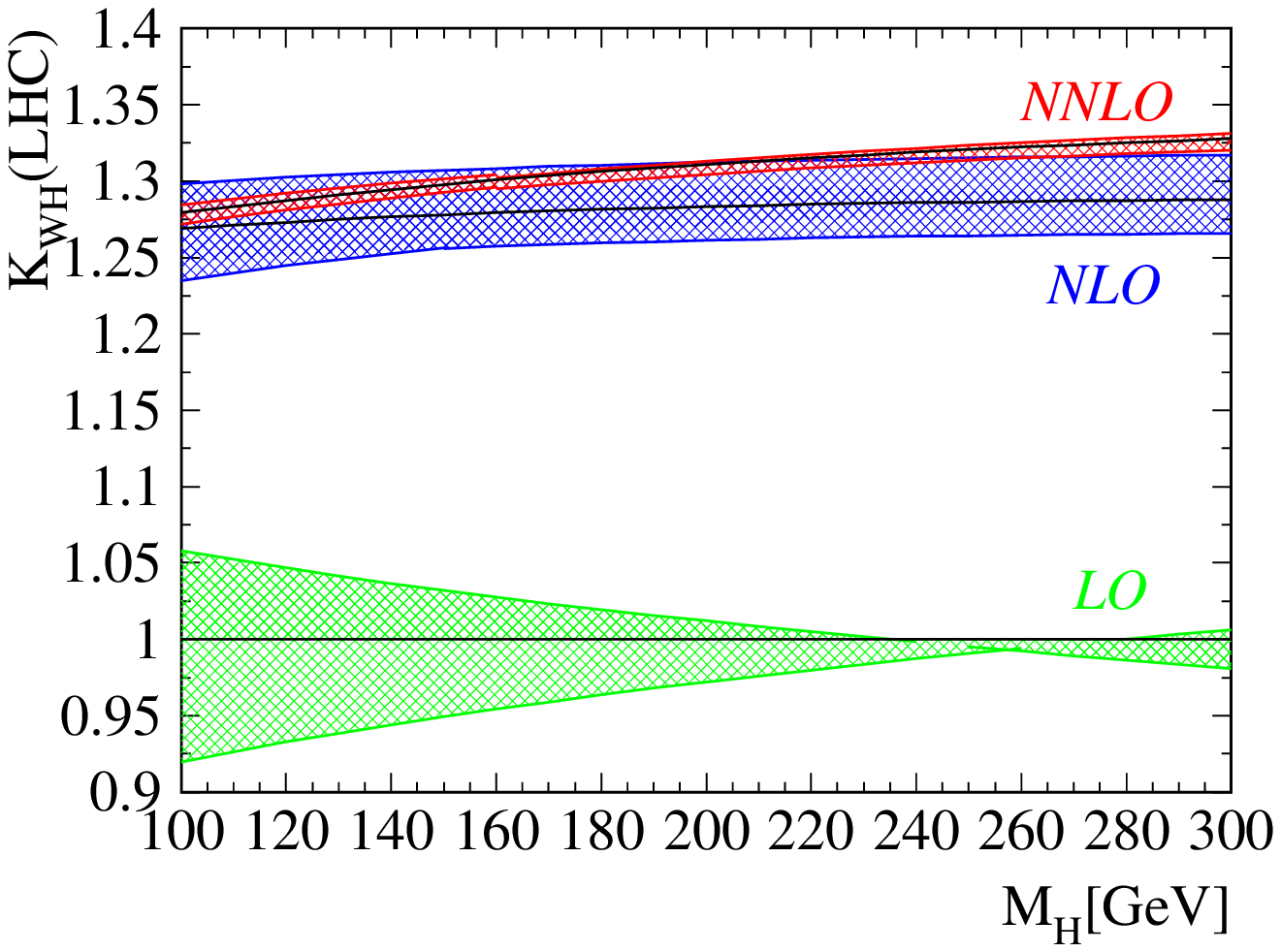}}
\put( 6.0,-5.7){\includegraphics{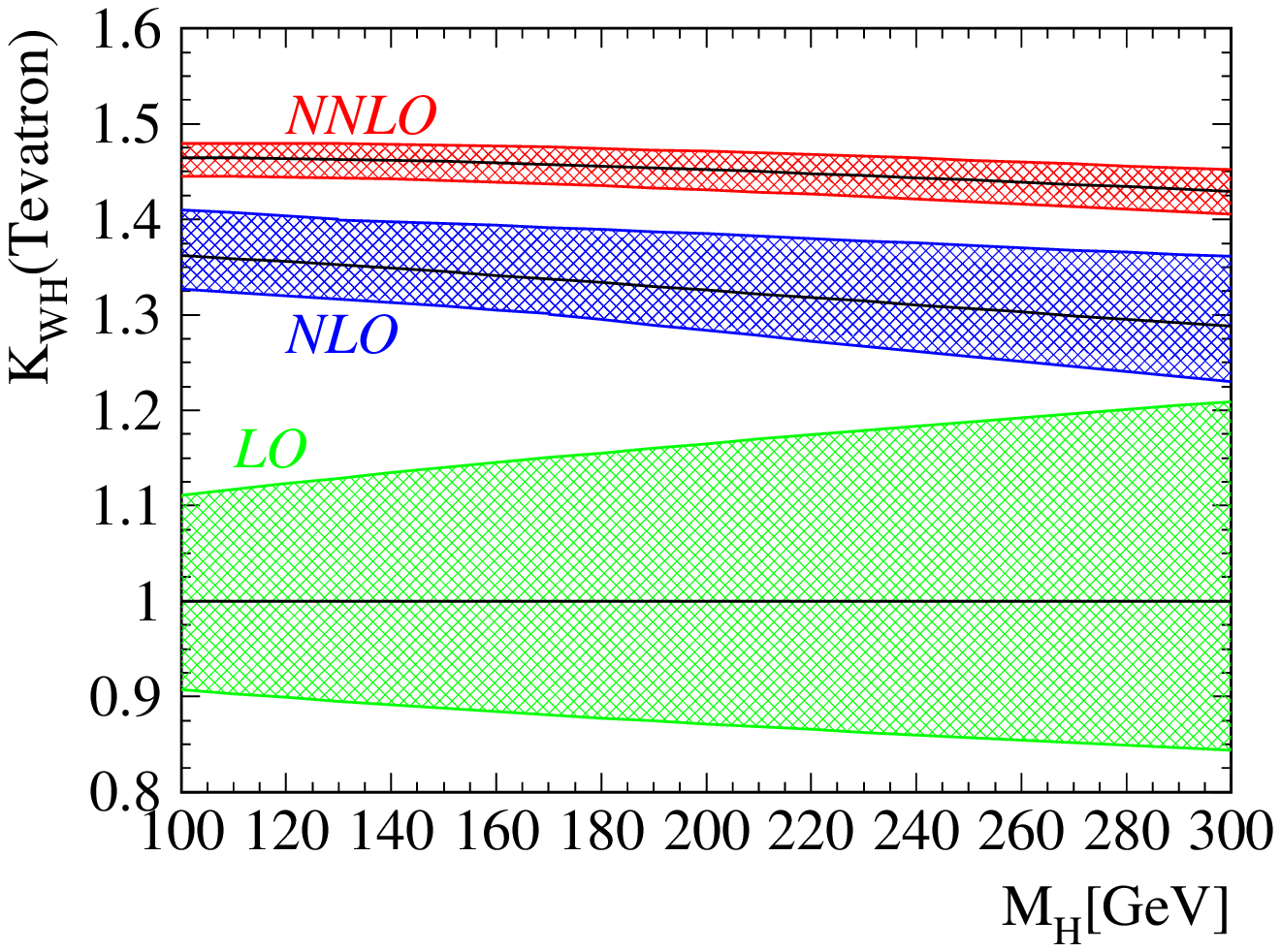}}
\end{picture} }
\caption{QCD $K$-factors for $WH$ production 
(i.e.\ from the sum of $W^+H$ and $W^-H$ cross sections) at the LHC
(l.h.s.)  and the Tevatron (r.h.s.). The bands represent the spread of
the cross section when the renormalization and factorization scales
are varied in the range $\frac{1}{3}M_{VH} \leq \mu_R\, (\mu_F) \leq
3M_{VH}$, the other scale being fixed at $\mu_F (\mu_R)= M_{VH}$.
(Taken from \citere{Brein:2003wg}.)}
\label{fig:Kfact}
\end{center}
\end{figure}

The $K$-factors at NLO and NNLO are shown in \reffi{fig:Kfact} 
(solid black lines)
for the LHC and the Tevatron  as a function of the Higgs  mass
$M_H$ for the process $p\bar p/pp \to WH+X$; they are practically the same 
for the process $p\bar p/pp \to ZH+X$ when the
contribution of the $gg \to ZH$ component is not included. 
Inclusion of this contribution adds substantially to the uncertainty 
of the NNLO prediction for $ZH$ production. This is because
$gg \to ZH$ appears at $\cal O(\alpha_{\rm s}^{\rm 2})$ in LO.

The  scales have
been fixed to $\mu_F=\mu_R=M_{VH}$, and the MRST sets of PDFs for each
perturbative order (including the NNLO PDFs of \citere{Martin:2002dr})
are used in a consistent manner.

The NLO $K$-factor is practically constant at the LHC, increasing only from
$K_{\rm NLO}=1.27$ for $M_H=110$ GeV to $K_{\rm NLO}=1.29$ for $M_H=300$ GeV.
The NNLO contributions increase the $K$-factor by a mere 1\% for the low $M_H$
value and by 3.5\% for the high value. At the Tevatron, the NLO $K$-factor is
somewhat higher than at the LHC, enhancing the cross section between  $K_{\rm
NLO}=1.35$ for $M_H=110$ GeV and $K_{\rm NLO}=1.3$ for $M_H=300$ GeV with a
monotonic decrease.  The NNLO corrections increase the $K$-factor uniformly by
about 10\%. Thus, these NNLO corrections are more important at the Tevatron
than at the LHC. 

The bands around the $K$-factors represent the cross section uncertainty due
to the variation of either the renormalization or factorization scale from
$\frac{1}{3} M_{VH} \leq \mu_F \, (\mu_R) \leq 3M_{VH}$, with the other scale
fixed at $\mu_R \, (\mu_F) = M_{VH}$; the
normalization is provided by the production cross section evaluated at scales
$\mu_F=\mu_R=M_{VH}$. As can be seen, except from the accidental cancellation
of the scale dependence of the LO cross section at the LHC, the decrease of the
scale variation is strong when going from LO to NLO and then to NNLO. For
$M_H=120$ GeV, the uncertainty from the scale choice at the LHC drops from 10\%
at LO, to 5\% at NLO, and to 2\% at NNLO. At the Tevatron and for the same
Higgs boson mass, the scale uncertainty drops from 20\% at LO, to 7\% at NLO,
and to 3\% at NNLO. If this variation of the cross section with the two scales
is taken as an indication of the uncertainties due to the not yet calculated
higher-order corrections, one concludes that once the NNLO QCD contributions are
included in the prediction, the QCD corrections to the cross section 
for the $p\bar p/pp \to VH+X$ process are known at the rather accurate level 
of 2 to 3\% relative to the LO.

\section{ELECTROWEAK CORRECTIONS}
\label{se:EW}

The calculation of the electroweak ${\cal O}(\alpha)$ corrections,
which employs established standard techniques, is described in
detail in \citere{Ciccolini:2003jy}. The virtual one-loop corrections
involve a few hundred diagrams, including self-energy, vertex, and
box corrections. In order to obtain IR-finite corrections, real-photonic 
bremsstrahlung has to be taken into account.
In spite of being IR finite, the ${\cal O}(\alpha)$ corrections
involve logarithms of the initial-state quark masses which are due to
collinear photon emission. These mass singularities are absorbed
into the PDFs in exactly the same way as in QCD, viz.\
by $\overline{\mbox{MS}}$ factorization.
As a matter of fact, this requires also the
inclusion of the corresponding ${\cal O}(\alpha)$ corrections into the
DGLAP evolution of these distributions and into their fit to
experimental data. At present, this full incorporation of ${\cal
O}(\alpha)$ effects in the determination of the quark distributions
has not been performed yet. However, an approximate inclusion of the
${\cal O}(\alpha)$ corrections to the DGLAP evolution shows
\cite{Kripfganz:1988bd} that the impact of these corrections on the
quark distributions in the $\overline{\mbox{MS}}$ factorization scheme
is well below 1\%, at least in the $x$ range that is relevant for
associated $VH$ production at the Tevatron and the LHC.  
This is also supported by a recent analysis of the MRST collaboration
\cite{Stirling} who took into account the ${\cal O}(\alpha)$ effects
to the DGLAP equations.

The size of the ${\cal O}(\alpha)$ corrections depends on the
employed input-parameter scheme for the coupling $\alpha$. 
This coupling can, for instance, be derived from
the fine-structure constant $\alpha(0)$, from the effective
running QED coupling $\alpha(M_Z^2)$ at the Z~resonance, or from
the Fermi constant $\GF$ via $\alpha_{\GF}=\sqrt{2}\GF M_W^2\sw^2/\pi$.
The corresponding schemes are known as $\alpha(0)$-, $\alpha(M_Z^2)$-,
and $\GF$-scheme, respectively.
In contrast to the $\alpha(0)$-scheme, where the ${\cal O}(\alpha)$ 
corrections are sensitive to the non-perturbative regime of the
hadronic vacuum polarization, in the $\alpha(M_Z^2)$- and $\GF$-schemes
these effects are absorbed into the coupling constant $\alpha$.
In the $\GF$-scheme large renormalization effects
induced by the $\rho$-parameter are absorbed in addition
via $\alpha_{\GF}$.
Thus, the $\GF$-scheme is preferable over the two other schemes
(at least over the $\alpha(0)$-scheme).

\begin{figure}
\begin{center}
\mbox{
\epsfig{file=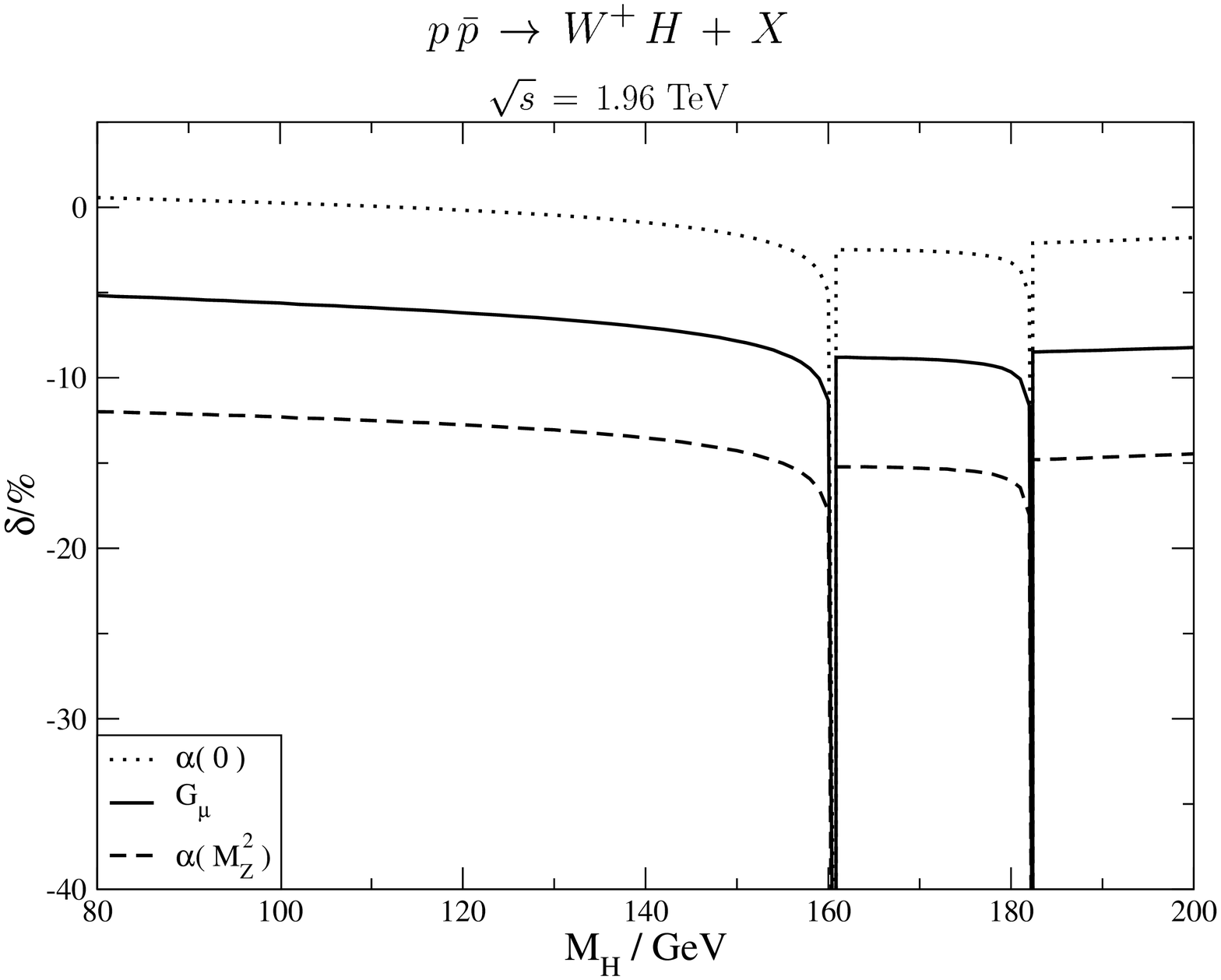,%
        bbllx=35pt,bblly=50pt,bburx=719pt,bbury=582pt,scale=0.33}
\hspace{.5em}
\epsfig{file=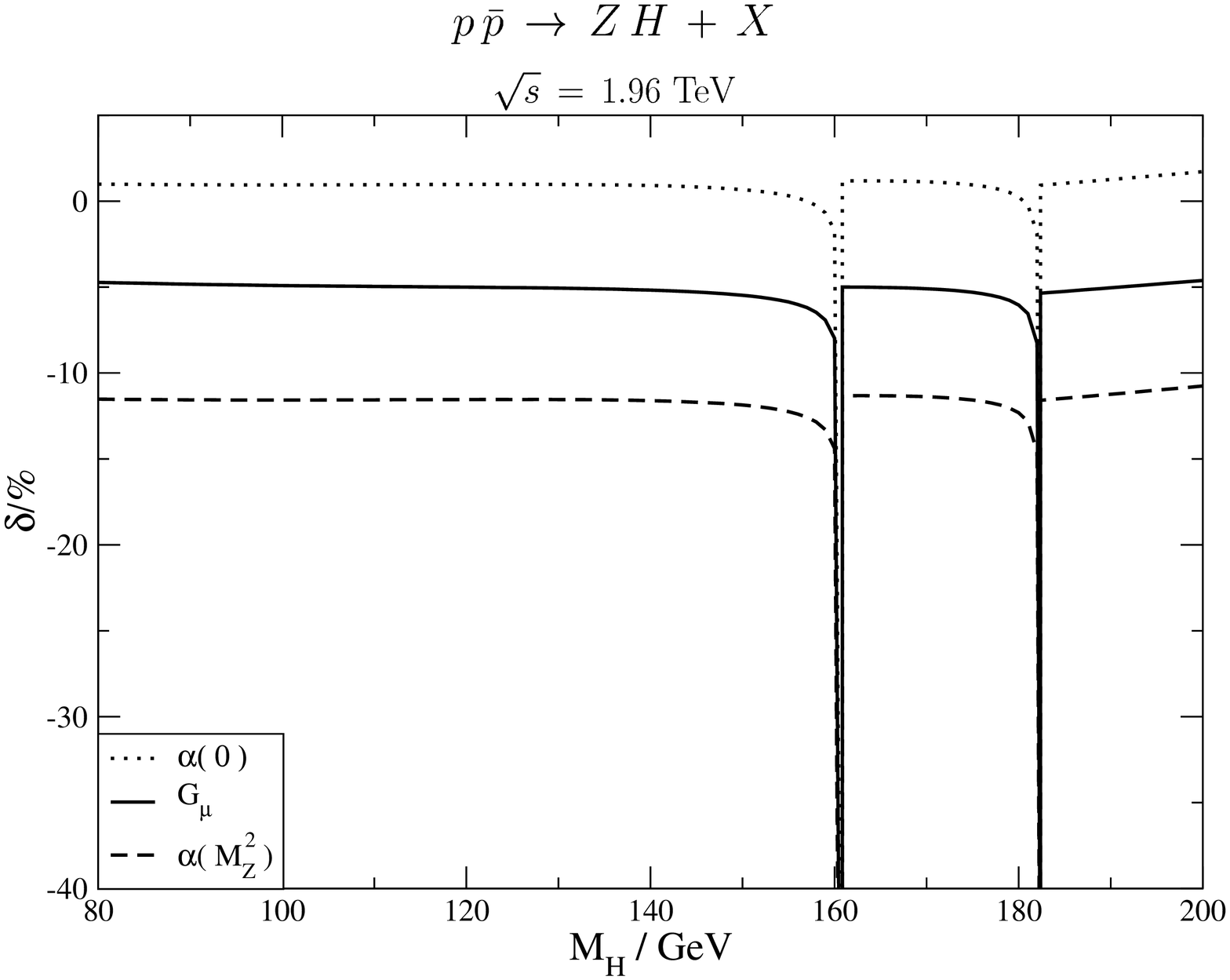,%
        bbllx=35pt,bblly=50pt,bburx=719pt,bbury=582pt,scale=0.33}
}
\vspace{.0em}
\caption{Relative electroweak correction $\delta$ as
 a function of $M_H$ for the total cross section of
 $p\bar{p}\to W^+H+X$ (l.h.s.) and $p\bar{p}\to ZH+X$ (r.h.s.)
at the Tevatron in various input-parameter schemes. 
(Taken from \citere{Ciccolini:2003jy}.)}
\label{fig:tevvh}
\end{center}
\end{figure}

Figure~\ref{fig:tevvh} shows the relative size of the ${\cal
  O}(\alpha)$ corrections as a function of the Higgs-boson mass for
$p\bar p \to W^+ H + X$ and $p\bar p \to ZH + X$ at the Tevatron. The
numerical results have been obtained using the
CTEQ6L1~\cite{Pumplin:2002vw} parton distribution function, but the
dependence of the relative electroweak correction $\delta$ displayed
in Fig.~\ref{fig:tevvh} on the PDF is insignificant.
Results are presented for the three
different input-parameter schemes. The corrections in the $\GF$- and
$\alpha(M_Z^2)$-schemes are significant and reduce the cross section
by 5--9\% and by 10--15\%, respectively. The corrections in the
$\alpha(0)$-scheme differ from those in the $\GF$-scheme by $2\Delta
r\approx 6\%$ and from those in the $\alpha(M_Z^2)$-scheme by
$2\Delta\alpha(M_Z^2)\approx 12\%$. 
The quantities $\Delta r$ and $\Delta\alpha(M_Z^2)$ denote, respectively, the
radiative corrections to muon decay and the correction describing
the running of $\alpha(Q^2)$ from $Q=0$ to $M_Z$
(see \citere{Ciccolini:2003jy} for details).
The fact that the relative
corrections in the $\alpha(0)$-scheme are rather small results from
accidental cancellations between the running of the electromagnetic
coupling, which leads to a contribution of about
$2\Delta\alpha(M_Z^2)\approx +12\%$, and other (negative) corrections
of non-universal origin.  Thus, corrections beyond ${\cal O}(\alpha)$
in the $\alpha(0)$-scheme cannot be expected to be suppressed as well.
In all schemes, the size of the corrections does not depend strongly
on the Higgs-boson mass.

For the LHC the corrections are similar in size
to those at the Tevatron and reduce the cross section by 5--10\% in
the $\GF$-scheme and by 12--17\% in the $\alpha(M_Z^2)$-scheme
(see Figs.~13 and 14 in \citere{Ciccolini:2003jy}).

In \citere{Ciccolini:2003jy} the origin of the electroweak corrections
was further explored by separating gauge-invariant building blocks.
It turns out that fermionic contributions (comprising all diagrams
with closed fermion loops) and remaining bosonic corrections partly
compensate each other, but the bosonic corrections are dominant.  The
major part of the corrections is of non-universal origin, i.e.\ the
bulk of the corrections is not due to coupling modifications, photon
radiation, or other universal effects.

\begin{figure}
\begin{center}
{ \unitlength 1cm
\begin{picture}(15.5,8.5)
\put(-2.2,-2){\includegraphics{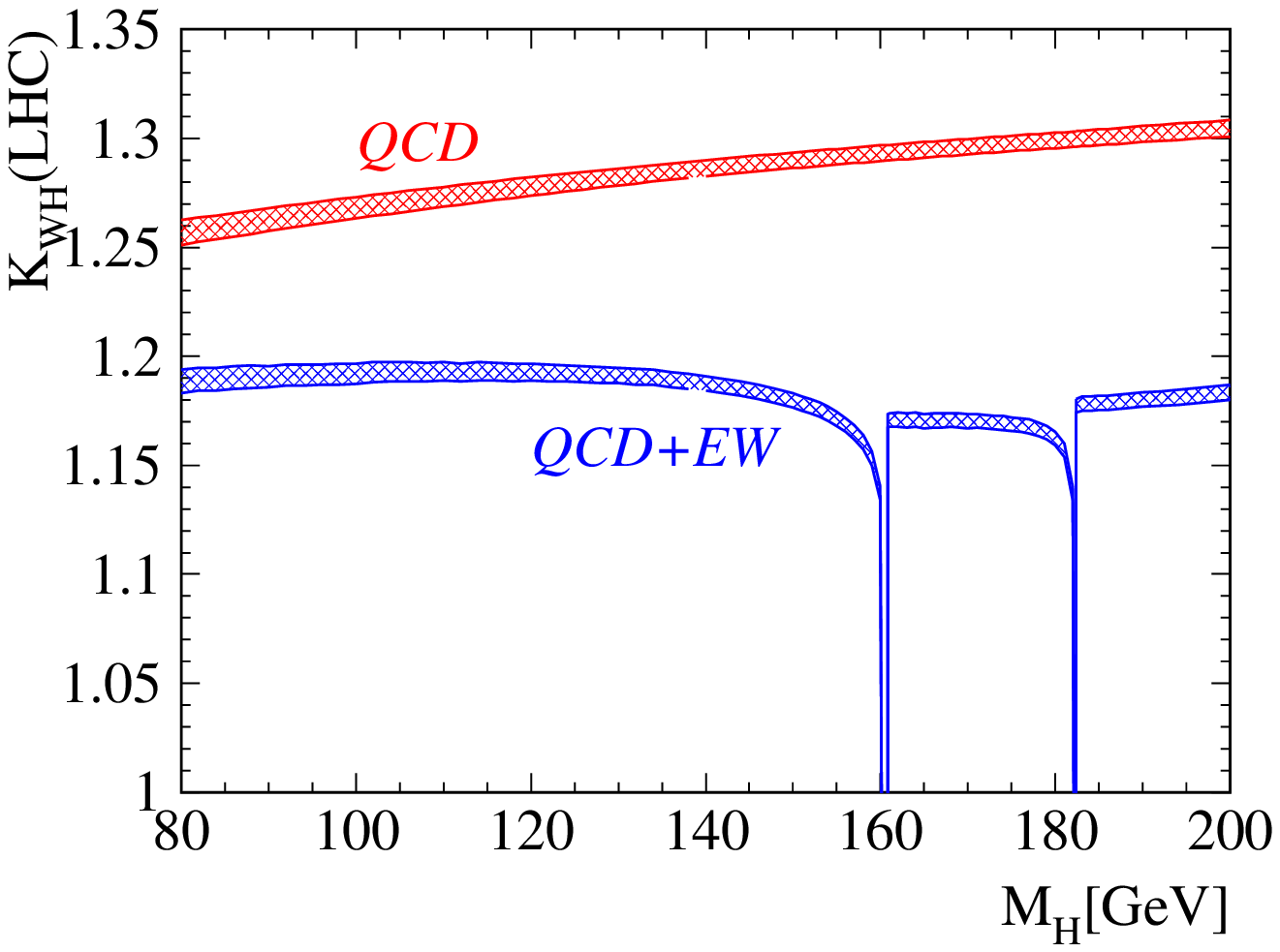}}
\put( 6.0,-2){\includegraphics{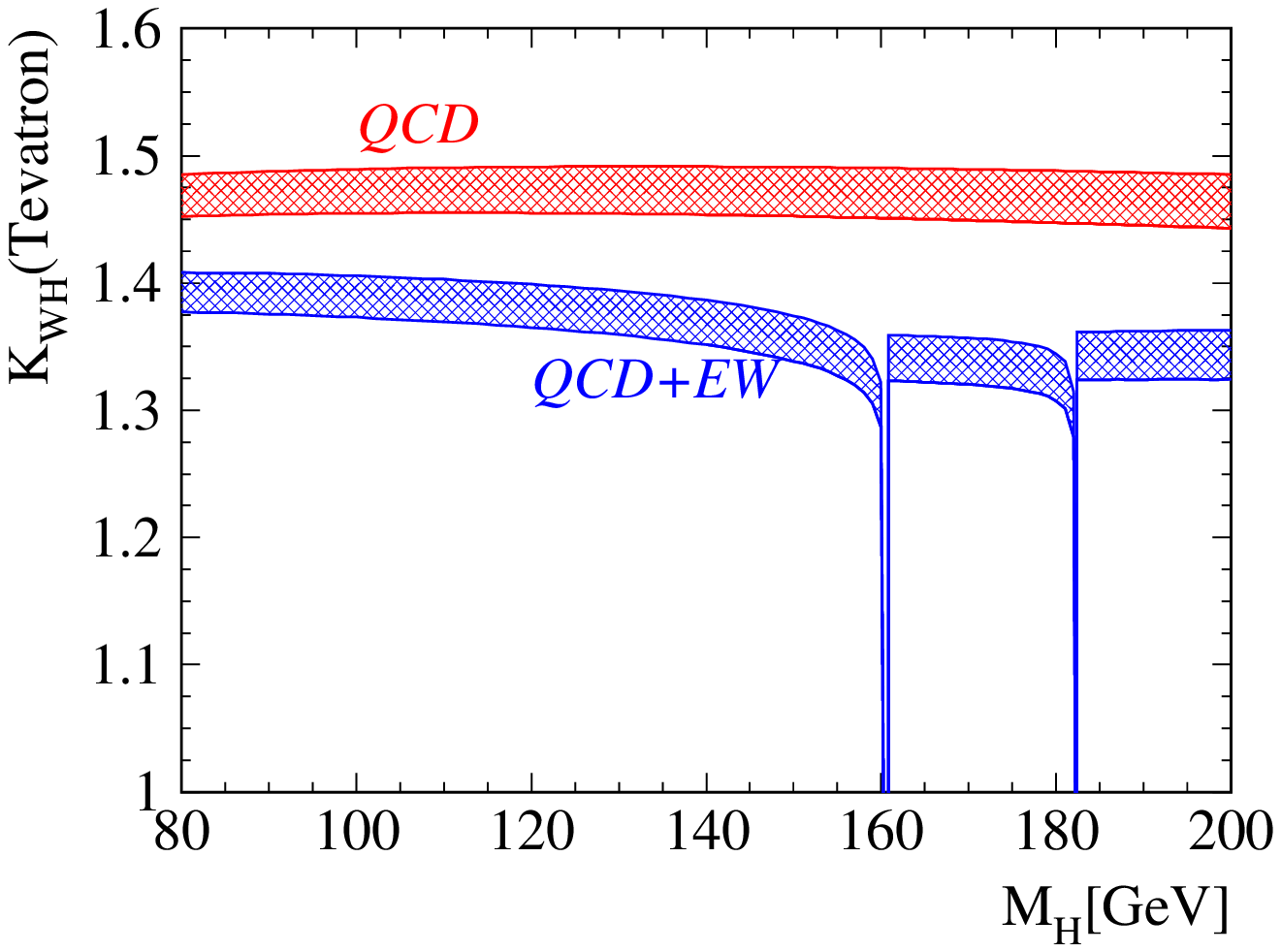}}
\put(-2.2,-8.5){\includegraphics{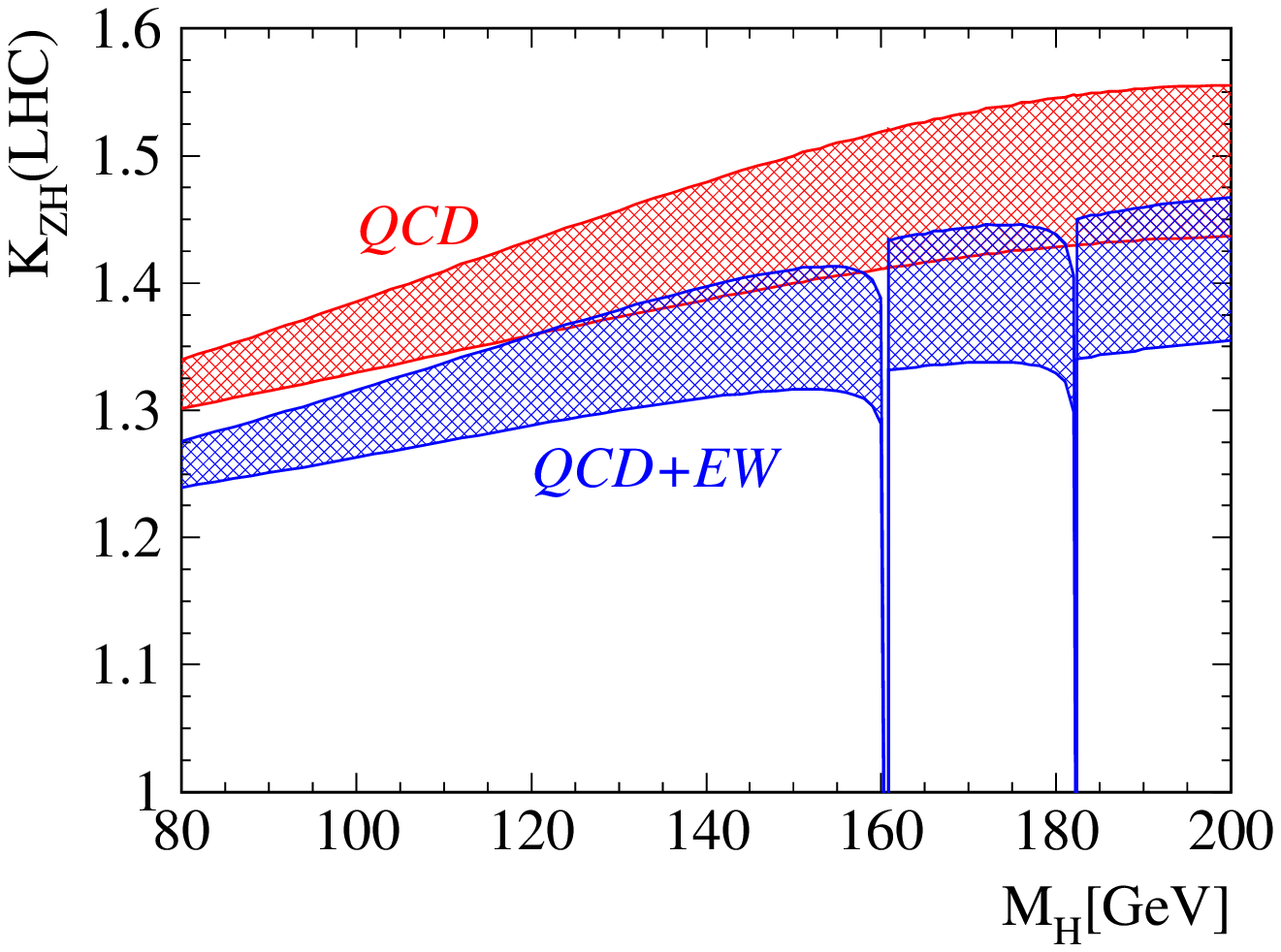}}
\put( 6.0,-8.5){\includegraphics{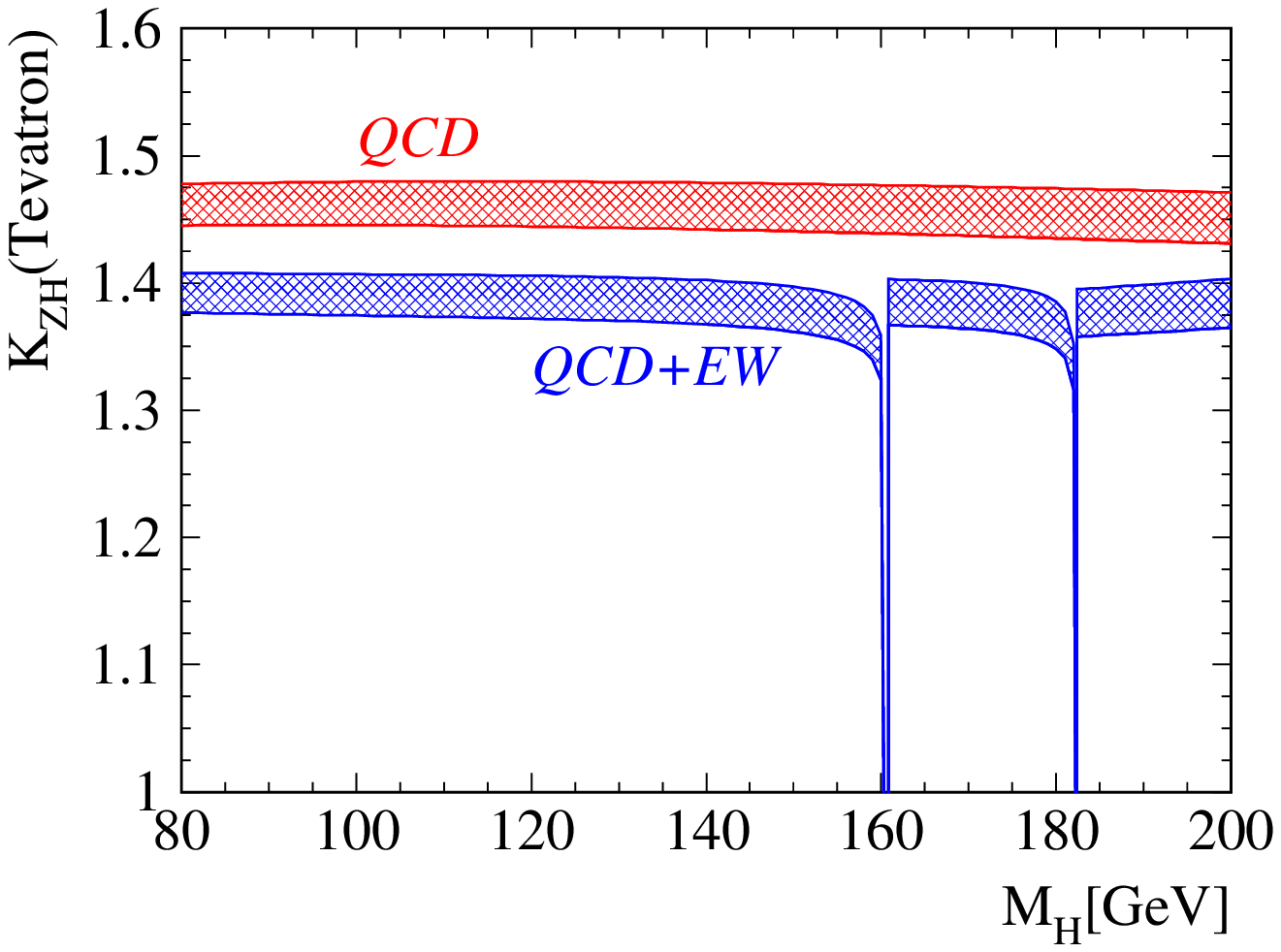}}
\end{picture} }
\vspace*{30mm}
\caption{
\label{fig:Kfactor}
$K$-factors for $WH$ production and $ZH$ production at the LHC
(l.h.s.) and the Tevatron (r.h.s.) after inclusion of the NNLO QCD and
electroweak ${\cal O}(\alpha)$ corrections. Theoretical errors as
described in Figure~\ref{fig:Kfact}.}
\end{center}
\end{figure}

Figure~\ref{fig:Kfactor} shows the $K$-factor after inclusion of both
the NNLO QCD and the ${\cal O}(\alpha)$ electroweak corrections for
$p\bar p/pp \to W H + X$ and $p\bar p/pp \to ZH + X$ at the Tevatron
and the LHC.  The larger uncertainty band for the $ZH$ production
process at the LHC is due to the contribution of $gg \to HZ$.

\section{CROSS-SECTION PREDICTIONS}
\label{se:numres}

\begin{figure}
\begin{center}
{ \unitlength 1cm
\begin{picture}(15.5,7)
\put(-4.4,-8.5){\includegraphics{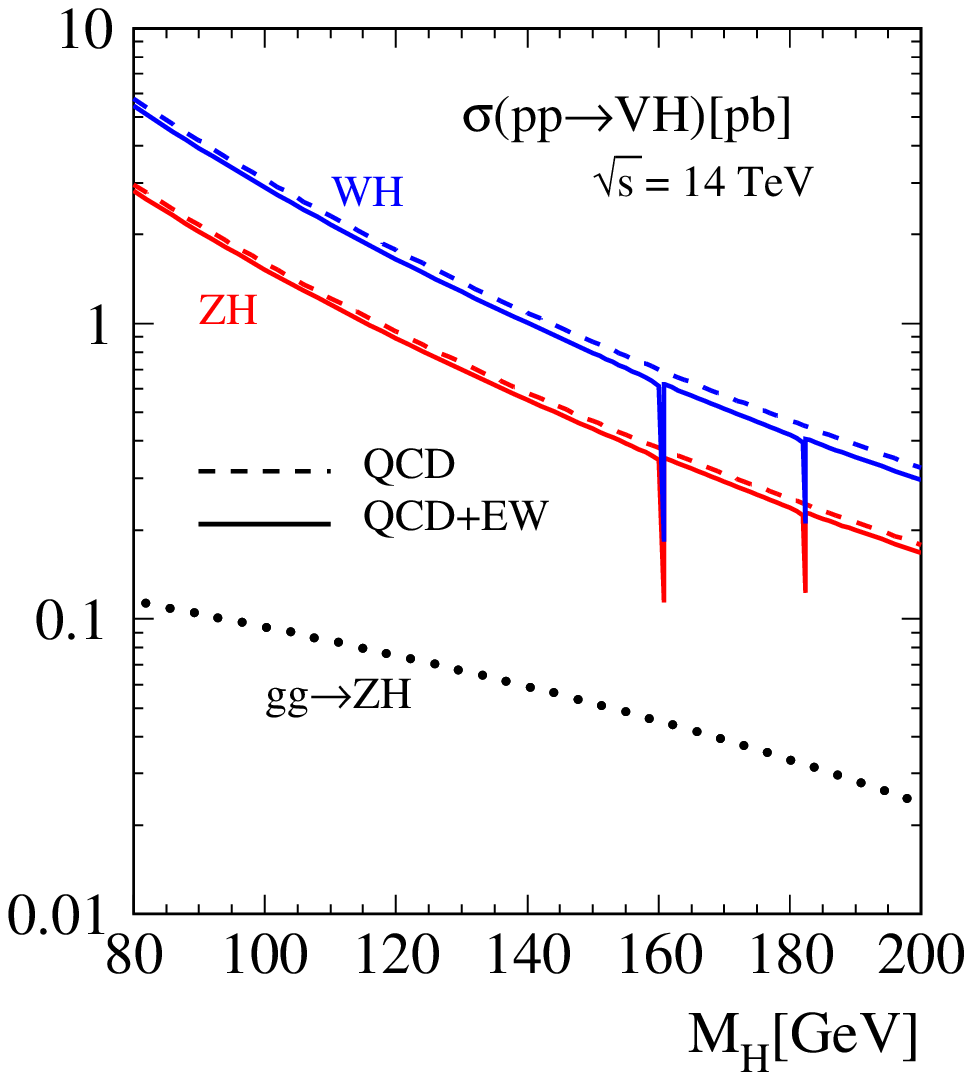}}
\put( 3.8,-8.5){\includegraphics{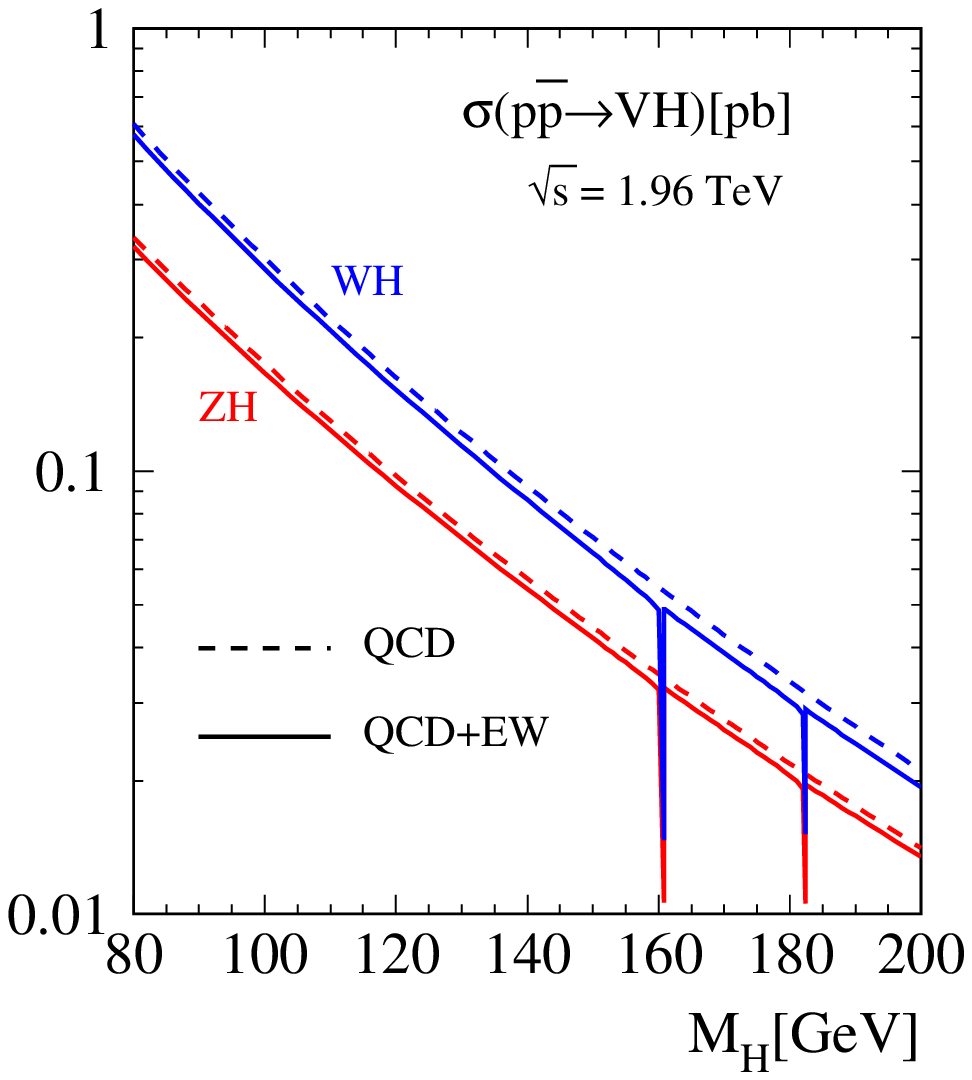}}
\end{picture} }
\vspace*{15mm}
\caption{
\label{fig:xsection}
Cross-section predictions (in the $\GF$-scheme)
for $WH$ and $ZH$ production 
at the LHC (l.h.s.) and the Tevatron (r.h.s.), including NNLO QCD
and electroweak ${\cal O}(\alpha)$ corrections.}
\end{center}
\end{figure}

Figure~\ref{fig:xsection} shows the predictions for the cross sections
of $WH$ and $ZH$ production at the LHC and the Tevatron, 
including the NNLO QCD and electroweak ${\cal O}(\alpha)$ corrections
as discussed in the previous sections.
At the LHC the process $gg\to ZH$ adds about 10\% to the $ZH$
production cross section, which is due to the large gluon flux;
at the Tevatron this contribution is negligible.

Finally, we briefly summarize the discussion \cite{Ciccolini:2003jy}
of the uncertainty in the cross-section predictions due to the error
in the parametrization of the parton densities (see also
\cite{Djouadi:2003jg}).  To this end the NLO cross section evaluated
using the default CTEQ6~\cite{Pumplin:2002vw} parametrization with the
cross section evaluated using the MRST2001~\cite{Martin:2002aw}
parametrization are compared. The results are collected in
\reftas{tab:pdfvhtev} and \ref{tab:pdfvhlhc}. Both the CTEQ and MRST
\begin{table}
\caption{\label{tab:pdfvhtev} Total cross sections (in fb) at
 the Tevatron ($\sqrt{s}=1.96\;\mathrm{TeV}$) including NLO QCD and
 electroweak corrections in the $\GF$-scheme for different sets of
 PDFs. The results include an estimate of the uncertainty due to the
 parametrization of the PDFs as obtained with the
 CTEQ6~\cite{Pumplin:2002vw} and MRST2001~\cite{Martin:2002aw}
 eigenvector sets. The renormalization and factorization scales have
 been set to the invariant mass of the Higgs--vector-boson pair, $\mu
 = \mu_0 = M_{VH}$. (Taken from \citere{Ciccolini:2003jy}.)}
\vspace{.5em}
\centerline{
\begin{tabular}{|c||c|c||c|c|}
\hline 
& \multicolumn{2}{c||}{$p\bar{p} \to WH+X$}
& \multicolumn{2}{c|}{$p\bar{p} \to ZH+X$}
\\
\hline 
$M_H/\mathrm{GeV}$ &
CTEQ6M~\cite{Pumplin:2002vw} & 
MRST2001~\cite{Martin:2002aw} &
CTEQ6M~\cite{Pumplin:2002vw} & 
MRST2001~\cite{Martin:2002aw} \\ \hline \hline
100.00  & 268.5(1)  $\pm$ 11    & 269.8(1) $\pm$ 5.2  &
          158.9(1)  $\pm$ 6.4   & 159.6(1) $\pm$ 2.0  \\ \hline
120.00  & 143.6(1)  $\pm$ 6.0   & 143.7(1) $\pm$ 3.0  &
           88.20(1) $\pm$ 3.6   & 88.40(1) $\pm$ 1.1  \\ \hline
140.00  &  80.92(1) $\pm$ 3.5   & 80.65(1) $\pm$ 1.8  &
           51.48(1) $\pm$ 2.1   & 51.51(1) $\pm$ 0.66 \\ \hline
170.00  &  36.79(1) $\pm$ 1.7   & 36.44(1) $\pm$ 0.91 &
           24.72(1) $\pm$ 1.0   & 24.69(1) $\pm$ 0.33 \\ \hline
190.00  &  22.94(1) $\pm$ 1.1   & 22.62(1) $\pm$ 0.60 &
           15.73(1) $\pm$ 0.68  & 15.68(1) $\pm$ 0.21 \\ \hline
\end{tabular}}
\vspace{2em}
\caption{\label{tab:pdfvhlhc} Same as in \refta{tab:pdfvhtev},
but for the LHC ($\sqrt{s}=14\;\mathrm{TeV}$)
(Taken from \citere{Ciccolini:2003jy}.)}
\vspace{.5em}
\centerline{
\begin{tabular}{|c||c|c||c|c|}
\hline 
& \multicolumn{2}{c||}{$pp \to WH+X$}
& \multicolumn{2}{c|}{$pp \to ZH+X$}
\\
\hline 
$M_H/\mathrm{GeV}$ 
& CTEQ6M~\cite{Pumplin:2002vw} & MRST2001~\cite{Martin:2002aw} 
& CTEQ6M~\cite{Pumplin:2002vw} & MRST2001~\cite{Martin:2002aw} 
\\ \hline \hline
100.00  & 2859(1)  $\pm$  96   & 2910(1)  $\pm$ 35  &
          1539(1)  $\pm$  51   & 1583(1)  $\pm$ 19  \\ \hline
120.00  & 1633(1)  $\pm$  55   & 1664(1)  $\pm$ 21  &
          895(3)   $\pm$  30   & 9217(3)  $\pm$ 11  \\ \hline
140.00  & 989(3)   $\pm$  34   & 1010(1)  $\pm$ 12  &
          551(2)   $\pm$  19   & 568.1(2) $\pm$ 6.7 \\ \hline
170.00  & 508(1)   $\pm$  18   & 519.3(1) $\pm$ 6.3 &
          290(1)   $\pm$  10   & 299.4(1) $\pm$ 3.6 \\ \hline
190.00  & 347(1)   $\pm$  12   & 354.7(2) $\pm$ 4.3 &
          197.8(1) $\pm$   6.9 & 204.5(1) $\pm$ 2.5 \\ \hline
\end{tabular}}
\end{table}
parametrizations include parton-distribution-error packages which
provide a quantitative estimate of the corresponding uncertainties in
the cross sections.%
\footnote{In addition, the MRST~\cite{Martin:2001es} parametrization allows to
study the uncertainty of the NLO cross section due to the variation of
$\alpha_{\rm s}$. For associated $WH$ and $ZH$ hadroproduction, the
sensitivity of the theoretical prediction to the variation of
$\alpha_{\rm s}$ ($\alpha_{\rm s}(M_Z^2) = 0.119\pm 0.02$) turns out
to be below $2\%$.}  Using the parton-distribution-error packages and
comparing the CTEQ and MRST2001 parametrizations, we find that the
uncertainty in predicting the $WH$ and $ZH$ production processes 
at the Tevatron and the LHC due to the
parametrization of the parton densities is less than approximately
$5\%$.

\section{CONCLUSIONS}
\label{se:concl}

After the inclusion of QCD corrections up to NNLO and of the
electroweak ${\cal O}(\alpha)$ corrections, the cross-section
predictions for $WH$ and $ZH$ production are by now the most
precise for Higgs production at hadron colliders. 
The remaining uncertainties
should be dominated by renormalization and factorization scale
dependences and uncertainties in the parton distribution
functions, which are of the order of 3\% and 5\%, respectively.
These uncertainties may be reduced by forming the ratios
of the associated Higgs-production cross section with the
corresponding Drell-Yan-like W- and Z-boson production channels,
i.e.\ by inspecting $\sigma_{p\bar{p}/pp\to VH+X}/\sigma_{p\bar{p}/pp\to
V+X}$, rendering their measurements particularly interesting 
at the Tevatron and/or the LHC.

\vskip1cm
\noindent

\section*{ACKNOWLEDGEMENTS}
We would like to thank the organizers of the Les Houches workshop for
their invitation and hospitality. M.~Kr\"amer would like to thank the
DESY Theory Group for their hospitality and financial support. This
work has been supported in part by the European Union under contract
HPRN-CT-2000-00149. M.~L.~Ciccolini is partially supported by ORS
Award ORS/2001014035. R.~Harlander is supported by DFG, contract HA 
2990/2-1.


\begin{thebibliography}{99}

\bibitem{Carena:2000yx}
Report of the Tevatron Higgs working group,
M.~Carena, J.~S.~Conway, H.~E.~Haber, J.~D.~Hobbs  {\it et al.}
[hep-ph/0010338].

\bibitem{Stange:ya}
A.~Stange, W.~J.~Marciano and S.~Willenbrock,
Phys.\ Rev.\ D {\bf 49} (1994) 1354
[hep-ph/9309294] and
Phys.\ Rev.\ D {\bf 50} (1994) 4491
[hep-ph/9404247].

\bibitem{atlas_cms_tdrs}
ATLAS Collaboration, Technical Design Report, Vols. 1 and 2, CERN--LHCC--99--14
and CERN--LHCC--99--15; \\
CMS Collaboration, Technical Proposal, CERN--LHCC--94--38; \\
A.~Djouadi {\it et al.},
``The Higgs working group: Summary report'',
proceedings of the Workshop On Physics At TeV Colliders, Les Houches, 
France, 1999 [hep-ph/0002258];\\
D.~Cavalli {\it et al.},
``The Higgs working group: Summary report'', 
proceedings of the Workshop On Physics At TeV Colliders, Les Houches, 
France, 2001 [hep-ph/0203056].

\bibitem{Glashow:ab}
S.~L.~Glashow, D.~V.~Nanopoulos and A.~Yildiz,
Phys.\ Rev.\ D {\bf 18} (1978) 1724;\\
Z.~Kunszt, Z.~Tr\'ocs\'any and W.~J.~Stirling,
Phys.\ Lett.\ B {\bf 271} (1991) 247.

\bibitem{Han:1991ia}
T.~Han and S.~Willenbrock,
Phys.\ Lett.\ B {\bf 273}, 167 (1991);\\
J.~Ohnemus and W.~J.~Stirling,
Phys.\ Rev.\ D {\bf 47} (1993) 2722;\\
H.~Baer, B.~Bailey and J.~F.~Owens,
Phys.\ Rev.\ D {\bf 47} (1993) 2730;\\
S.~Mrenna and C.~P.~Yuan,
Phys.\ Lett.\ B {\bf 416} (1998) 200
[hep-ph/9703224];\\
M.~Spira,
Fortsch.\ Phys.\  {\bf 46} (1998) 203
[hep-ph/9705337];\\
A.~Djouadi and M.~Spira,
Phys.\ Rev.\ D {\bf 62} (2000) 014004 
[hep-ph/9912476].

\bibitem{Dicus:1985wx}
D.~A.~Dicus and S.~S.~Willenbrock,
Phys.\ Rev.\ D {\bf 34} (1986) 148;\\
D.~A.~Dicus and C.~Kao,
Phys.\ Rev.\ D {\bf 38} (1988) 1008
[Erratum-ibid.\ D {\bf 42} (1990) 2412];\\
V.~D.~Barger, E.~W.~Glover, K.~Hikasa, W.~Y.~Keung, M.~G.~Olsson, C.~J.~Suchyta and X.~R.~Tata,
Phys.\ Rev.\ Lett.\  {\bf 57} (1986) 1672;\\
B.~A.~Kniehl,
Phys.\ Rev.\ D {\bf 42} (1990) 2253.

\bibitem{Brein:2003wg}
O.~Brein, A.~Djouadi and R.~Harlander,
Phys.\ Lett.\ B {\bf 579} (2004) 149
[hep-ph/0307206].

\bibitem{Ciccolini:2003jy}
M.~L.~Ciccolini, S.~Dittmaier and M.~Kr\"amer,
Phys.\ Rev.\ D {\bf 68} (2003) 073003
[hep-ph/0306234].

\bibitem{Hamberg:1990np}
R.~Hamberg, W.~L.~van Neerven and T.~Matsuura,
Nucl.\ Phys.\ B {\bf 359} (1991) 343
[Erratum-ibid.\ B {\bf 644} (2002) 403];\\
W.~L.~van Neerven and E.~B.~Zijlstra,
Nucl.\ Phys.\ B {\bf 382} (1992) 11.

\bibitem{Harlander:2002wh}
R.~V.~Harlander and W.~B.~Kilgore,
Phys.\ Rev.\ Lett.\  {\bf 88} (2002) 201801
[hep-ph/0201206].

\bibitem{Barger:1986jt}
V.~D.~Barger {\it et al.},
Phys.\ Rev.\ Lett.\  {\bf 57} (1986) 1672;\\
%
D.~A.~Dicus and C.~Kao,
Phys.\ Rev.\ D {\bf 38} (1988) 1008
[Erratum-ibid.\ D {\bf 42} (1990) 2412];\\
%
B.~A.~Kniehl,
Phys.\ Rev.\ D {\bf 42} (1990) 2253.

\bibitem{Martin:2002dr}
A.~D.~Martin, R.~G.~Roberts, W.~J.~Stirling and R.~S.~Thorne,
Phys.\ Lett.\ B {\bf 531} (2002) 216
[hep-ph/0201127].

\bibitem{Kripfganz:1988bd}
J.~Kripfganz and H.~Perlt,
Z.\ Phys.\ C {\bf 41} (1988) 319;\\
H.~Spiesberger,
Phys.\ Rev.\ D {\bf 52} (1995) 4936
[hep-ph/9412286].

\bibitem{Stirling}
W.~J.~Stirling, talk given at the workshop
{\it Electroweak Radiative Corrections to Hadronic Observables at
TeV Energies}, Durham, UK, September 2003.

\bibitem{Pumplin:2002vw}
J.~Pumplin, D.~R.~Stump, J.~Huston, H.~L.~Lai, P.~Nadolsky and W.~K.~Tung,
JHEP {\bf 0207} (2002) 012
[hep-ph/0201195].

\bibitem{Djouadi:2003jg}
A.~Djouadi and S.~Ferrag,
hep-ph/0310209.

\bibitem{Martin:2002aw}
A.~D.~Martin, R.~G.~Roberts, W.~J.~Stirling and R.~S.~Thorne,
Eur.\ Phys.\ J.\ C {\bf 28} (2003) 455
[hep-ph/0211080].

\bibitem{Martin:2001es}
A.~D.~Martin, R.~G.~Roberts, W.~J.~Stirling and R.~S.~Thorne,
Eur.\ Phys.\ J.\ C {\bf 23} (2002) 73
[hep-ph/0110215].

\end{thebibliography}
\end{document}